\documentstyle[12pt,a4]{article}
\begin{document}
\begin{center}
{\bf \Large{GENERALIZED KILLING EQUATIONS FOR \\
~\\                                                                  
SPINNING SPACES AND THE ROLE OF \\
~\\
KILLING-YANO TENSORS}}
\end{center}
\vskip5mm
\centerline{{\bf\large Mihai Visinescu\footnote
{E-mail address:~~~ MVISIN@THEOR1.IFA.RO}}}
\vskip5mm
\centerline{Department of Theoretical Physics}
\centerline{Institute of Atomic Physics, P.O.Box MG-6, Magurele,}
\centerline{Bucharest, Romania}                                      
\vskip5mm 
\nopagebreak \begin{abstract}
\noindent

The generalized Killing equations for the configuration space of
spinning particles (spinning space) are analysed. Solutions of
these equations are expressed in terms of Killing-Yano tensors.
In general the constants of motion can be seen as extensions of 
those from the scalar case or new ones depending on the
Grassmann-valued spin variables.
\end{abstract}
\vskip5mm
\par
Spinning particles, such as Dirac fermions, can be described by 
pseudo-classical mechanics models involving anticommuting c-numbers 
for the spin-degrees of freedom. The configuration space of spinning 
particles (spinning space) is an extension of an ordinary Riemannian 
manifold, parametrized by local coordinates {$\{$}$x^\mu${$\}$}, to 
a graded manifold parametrized by local coordinates {$\{$}$x^\mu, 
\psi^\mu${$\}$}, with the first set of variables being Grassmann-even 
( commuting ) and the second set Grassmann-odd (anticommuting) [1-3].

The equation of motion of a spinning particle on a geodesic is
derived from the action:
\begin{equation}
S=\int d\tau\left( \frac12 g_{\mu\nu}(x)\dot{x}^\mu \dot{x}^\nu + 
\frac{i}{2} 
g_{\mu\nu}(x)\psi^\mu\frac{D\psi^\nu}{D\tau}\right) .
\end{equation}

The corresponding world-line hamiltonian is given by:
\begin{equation}
H=\frac1 2 g^{\mu\nu}\Pi_\mu \Pi_\nu
\end{equation}
where $\Pi_\mu = g_{\mu\nu}\dot{x}^\nu$ is the covariant 
momentum. 

For any constant of motion ${\cal J}(x,\Pi,\psi)$, the bracket
with $H$ vanishes
\begin{equation}
\left\lbrace H,{\cal J} \right\rbrace = 0.
\end{equation}

If we expand ${\cal J}(x,\Pi,\psi)$ in a power series in the
canonical momentum 
\begin{equation}
{\cal J}=\sum_{n=0}^{\infty}\frac{1}{n!}{\cal J}^{(n)\mu_1
\dots\mu_n}(x,\psi) \Pi_{\mu_1}\dots\Pi_{\mu_n}
\end{equation}
then the bracket $\{ H , {\cal J}\}$ vanishes for
arbitrary $\Pi_\mu$ if and only if the components of ${\cal J}$ 
satisfy the generalized Killing equations [1] :
\begin{equation}
{\cal J}^{(n)}_{(\mu_1\dots\mu_n;\mu_{n+1})} + \frac{\partial
{\cal J}^{(n)}_{(\mu_1 \dots\mu_n}}{\partial \psi^\sigma}
\Gamma^\sigma_{\mu_{n+1})\lambda} \psi^\lambda = 
\frac{i}{2}\psi^\rho \psi^\sigma R_{\rho\sigma\nu(\mu_{n+1}}
{{\cal J}^{(n+1)\nu}}_{\mu_1 \dots \mu_n)}
\end{equation}
where the parentheses denote full symmetrization over the 
indices enclosed.

In general the symmetries of a spinning-particle model can be divided
into two classes.  First, there are four independent {\it generic}
symmetries  which exist in any theory [1-3] :

\begin{enumerate}
\item{Proper-time translations generated by the hamiltonian $H$ (2)}
\item{Supersymmetry generated by the supercharge
\begin{equation}
Q=\Pi_\mu\,\psi^\mu
\end{equation}}
\item{Chiral symmetry generated by the chiral charge
\begin{equation}
\Gamma_* = \frac{i^{[\frac{d}{2}]}}{d!}\sqrt{g}\epsilon_{\mu_1 
\dots \mu_d} \psi^{\mu_1} \dots \psi^{\mu_d}
\end{equation}}
\item{Dual supersymmetry, generated by the dual supercharge
\begin{equation}
Q^* = i\{ \Gamma_* , Q_0 \} = 
\frac{i^{[\frac{d}{2}]}}{(d-1)!}\sqrt{g}\epsilon_{\mu_1 \dots 
\mu_d} \Pi^{\mu_1}\psi^{\mu_2} \dots \psi^{\mu_d}
\end{equation}}
\end{enumerate}
where $d$ is the dimension of space-time.

The second kind of conserved quantities, called {\it
non-generic}, depend on the explicit form of the metric
$g_{\mu\nu}(x)$. In the recent literature there are exhibited the
constants of motion in the Schwarzschild [4], Taub-NUT [5-7],
Kerr-Newman [3] spinning spaces.

In what follows we shall deal with the {\it non-generic}
constants of motion in connection with the Killing equations (5)
looking for the general features of the solutions. In general
the constants of motion can seen as extensions of the constants from
the scalar case or new ones depending on the Grassmann-valued
spin variables {$\{$}$\psi^\mu${$\}$}.
 
Let us assume that the number of terms in the series (4) is
finite. That means that, for a given $n$, 
~${\cal J}^{(n+1)}_{(\mu_1 \dots\mu_{n+1})}$
vanishes and the last non-trivial equation from the system of
Killing equations (5) becomes homogeneous :
\begin{equation}
{\cal J}^{(n)}_{(\mu_1 \dots\mu_n;\mu_{n+1})}+
\frac{\partial {\cal J}^{(n)}_{(\mu_1 \dots\mu_n}}
{\partial \psi^\sigma}
\Gamma^\sigma_{\mu_{n+1})\lambda}\psi^\lambda = 0.
\end{equation}

In order to solve the system of coupled differential equations
(5) one starts with a ${\cal J}^{(n)}_{\mu_1 \dots
\mu_n}$ solution of the homogeneous equation (9).
This solution is introduced in the right-hand side (RHS) of 
the generalized Killing equation (5) for ${\cal J}^{(n-1)}_{\mu_1
\dots\mu_{n-1}}$ and the iteration is carried on to $n=0$.

In fact, for the bosonic sector, neglecting the Grassmann
variables $\{ \psi^\mu \}$, all the generalized Killing
equations (5) are homogeneous and decoupled. The first equation
shows that ${\cal J}^{(0)}$ is a trivial constant, the next one is
the equation for the Killing vectors and so on. In general, the
homogeneous equation (9) for a given $n$ in which all spin
degrees of freedom are neglected, defines a Killing tensor of
valence $n$
\begin{equation} 
{\cal J}^{(n)}_{\mu_1\dots\mu_n;\mu_{n+1}} = 0
\end{equation}
and 
\begin{equation}
{\cal J} ={\cal J}^{(n)}_{\mu_1
\dots\mu_n}\Pi^{\mu_1}\dots\Pi^{\mu_n}
\end{equation}
is a first integral of the geodesic equation [8].

For the spinning particles, even if one starts with a Killing
tensor of valence $n$, solution of eq.(10) in which all spin
degrees of freedom are neglected, the components 
${\cal J}^{(m)}_{\mu_1\dots\mu_m}~~(m<n)$ will receive a 
nontrivial spin contribution. 

Therefore the quantity (11) is no more conserved and
the actual constant of motion is
\begin{equation}
{\cal J}=\sum_{m=0}^{n}\frac{1}{m!}{\cal J}^{(m)\mu_1
\dots\mu_m} \Pi_{\mu_1}\dots\Pi_{\mu_m}
\end{equation}
in which ${\cal J}^{(m)\mu_1\dots\mu_m}$ with $m<n$ has a
nontrivial spin-dependent expression.

We shall illustrate the above construction with a few examples.
Since for $n=0$ eq. (10) is trivial, we shall consider the
next case, namely $n=1$. In this case eq. (10) is satisfied
by a Killing vector  $R_\mu$ 
\begin{equation}
 R_{(\mu;\nu)} = 0.
\end{equation}

Introducing this Killing vector in the RHS of the generalized
Killing equation (5) for $n=0$ one obtains for the
Killing scalar [7]
\begin{equation}
{\cal J}^{(0)} = \frac i2 R_{[\mu;\nu]} \psi^\mu \psi^\nu 
\end{equation}
where the square bracket denotes antisymmetrization with norm one.

A more involved example is given by a Killing tensor $K_{\mu\nu}$  
satisfying equation (10) for $n=2$:
\begin{equation}
K_{(\mu\nu;\lambda)} = 0.
\end{equation}

Unfortunately it is not possible to find a closed, analytic
expression for the spin corrections to the quantity (11) in
terms of the components of the Killing tensor $K_{\mu\nu}$ and
its derivatives. But assuming that the Killing tensor
$K_{\mu\nu}$ can be written as a symmetrized product of two
Killing-Yano tensors, the construction of the conserved quantity
(12) is feasible. We remind that a tensor  $f_{\mu_1 \dots\mu_r}$ 
is called Killing-Yano of valence $r$ if it is totally antisymmetric 
and satisfies the equation  [9]
\begin{equation}
f_{\mu_1 \dots\mu_{r-1}(\mu_{r};\lambda)} = 0.
\end{equation}

For the generality, let us assume that the Killing tensor can be
written as a symmetrized product of two different Killing-Yano 
tensors
\begin{equation}
 K^{\mu\nu}_{ij} = {1\over 2}(f^\mu_{i~\lambda} f^{\nu\lambda}_j +
f^\nu_{i~\lambda} f^{\mu\lambda}_j)
\end{equation}
where $f^{\mu\nu}_{i}$ is a Killing Yano tensor of type $i$ and
the Killing tensor has two additional indices $i,j$ to record
the fact that it is formed from two different Killing-Yano
tensors $(i\neq j)$.

Introducing the Killing tensor (17) in the RHS of eq.(5) for
$n=1$ we get a spin contribution to the Killing vector [3]:
\begin{equation}
I^{(1)\mu}_{ij} = {i\over 2}\Psi^\lambda \Psi^\sigma (f^\nu_{i~\sigma}
D_\nu f^\mu_{j~\lambda} + f^\nu_{j~\sigma} D_\nu f^\mu_{i~\lambda}
+ {1\over 2} f^{\mu\rho}_i c_{j\lambda\sigma\rho} +
 f^{\mu\rho}_j c_{i\lambda\sigma\rho} )
\end{equation}
and using this quantity in the RHS of eq.(5) for $n=0$ we get
for the Killing scalar
\begin{equation}
J^{(0)}_{ij} =-{1\over 4} \Psi^\lambda\Psi^\sigma\Psi^\rho\Psi^\tau
(R_{\mu\nu\lambda\sigma} f^\mu_{i~\rho} f^\nu_{j~\tau} + {1\over 2}
c^{~~~\pi}_{i\lambda\sigma} c_{j\rho\tau\pi})
\end{equation}
where the tensor $c_{i\mu\nu\lambda}$ is [3] :
\begin{equation}
c_{i\mu\nu\lambda} = -2 f_{i[\nu\lambda;\mu]}.
\end{equation}

Higher orders of the generalized equations (5) can be treated
similarly, but the corresponding expressions are quite involved.

In what follows we shall analyze the homogeneous equation (9)
looking for solutions depending on the Grassmann variables
{$\{$}$\psi^\mu${$\}$}. Even the lowest order equation with
$n=0$ has a nontrivial solution [7,10]
\begin{equation}
{\cal J}^{(0)} = \frac i4 f_{\mu\nu}\psi^\mu\psi^\nu
\end{equation}
where $f_{\mu\nu}$ is a Killing-Yano tensor
covariantly constant. Moreover ${\cal J}^{(0)}$ is a
separately conserved quantity.

Going to the next equation (9) with $n=1$, a natural solution
is: 
\begin{equation}
{\cal J}^{(1)}_\mu = R_\mu f_{\lambda\sigma}\psi^\lambda
\psi^\sigma
\end{equation}
where $R_\mu$ is a Killing vector and
again $f_{\lambda\sigma}$ is a Killing-Yano tensor covariantly
constant. Introducing this solution in the RHS of the eq. (5)
with $n=0$, after some calculations, we get for ${\cal J}^{(0)}$
[10]:
\begin{equation}
{\cal J}^{(0)} = \frac i2 R_{[\mu;\nu]}f_{\lambda\sigma}\psi^\mu 
\psi^\nu \psi^\lambda \psi^\sigma
\end{equation}

Combining eq.(22) and (23) with the aid of eqs.(12) 
we get a constant of motion which is peculiar to the
spinning case:
\begin{equation}
{\cal J} = f_{\mu\nu}\psi^\mu \psi^\nu \left( R_\lambda
\Pi^\lambda + \frac i2 R_{[\lambda;\sigma]}\psi^\lambda \psi^\sigma
\right).
\end{equation}
In fact this constant of motion is not completely new and it can
be expressed in terms of the quantities (14) and (21).

Another $\psi$-dependent solution of eq.(9) for $n=1$ can be
generated from a Killing-Yano tensor of valence $r$:
\begin{equation}
{\cal J}_{\mu_1}^{(1)} = f_{\mu_1 \mu_2\dots\mu_r}\psi^{\mu_2}
\dots \psi^{\mu_r}.
\end{equation}

Again introducing this quantity in the RHS of eq.(5) for $n=0$
we get for ${\cal J}^{(0)}$:
\begin{equation}
{\cal J}^{(0)} = \frac{i}{r+1}(-1)^{r+1} f_{[\mu_1\dots
\mu_r;\mu_{r+1}]} \psi^{\mu_1}\dots\psi^{\mu_{r+1}}
\end{equation}
and the constant of motion corresponding to these solutions
of the Killing equations is [10]:
\begin{equation}
Q_f = f_{\mu_1 \dots\mu_r}\Pi^{\mu_1}\psi^{\mu_2}\dots \psi^{\mu_r} 
+ \frac{i}{r+1}(-1)^{r+1}f_{[\mu_1 \dots
\mu_r;\mu_{r+1}]}\psi^{\mu_1}\dots \psi^{\mu_{r+1}}.
\end{equation}

Therefore the existence of a Killing-Yano tensor of valence $r$ 
is equivalent to the existence of a supersymmetry for the
spinning space with supercharge $Q_f$ which anticommutes with
$Q_0$. A similar result was obtained in ref.[11] in which it is
discussed the role of the generalized Killing-Yano tensors, with
the framework extended to include electromagnetic interactions.

The aim of this paper was to point out the important role of the
Killing-Yano tensors to generate solutions of the generalized
Killing equations. This aspect is closely connected with the
fact that the Killing-Yano tensors can be understood as objects
generating {\it non-generic} supersymmetries [3]. For the first
orders of eqs.(5) and (9) $(n\leq 2)$, which are usually encountered 
in theories of interest, we presented the complete form of the
solutions. With some ability, it is possible to investigate the
higher orders of the generalized Killing equations, but it seems 
that one cannot go much far with simple, transparent expressions. 
The extension of these results for the motion of spinning particles 
in spaces with torsion and/or in the presence of an electromagnetic 
field will be discussed elsewhere.

\subsection*{Acknowledgements}~~
The author wishes to thank the organizers of the Ahrenshoop
Symposium, Buckow 1996, for their kind invitation and financial
support that made possible his participation in this very
enjoyable event.
\end{document}